\documentclass[aps,pre]{revtex4}
\usepackage{graphicx}

\begin{document}

\title{Modulational Instabilities and Domain Walls in Coupled Discrete Nonlinear Schr\"odinger
Equations}
\author{Z. Rapti$^{1,2}$, A. Trombettoni$^3$, P.G. Kevrekidis$^{1,2}$, D.J.
Frantzeskakis$^4$, Boris A. Malomed$^5$ and A.R. Bishop$^2$}
\affiliation{$^1$Department of Mathematics and Statistics, University of Massachusetts,
Amherst MA 01003-4515, USA \\
$^2$ Center for Nonlinear Studies and Theoretical Division, Los Alamos
National Laboratory, Los Alamos, NM 87545, USA \\
$^{3}$ Istituto Nazionale per la Fisica della Materia, and Dipartimento di
Fisica, Universita' di Parma, parco Area delle Scienze 7A, I-43100 Parma,
Italy \\
$^{4}$ Department of Physics, University of Athens, Panepistimiopolis,
Zografos, Athens 15784, Greece \\
$^5$Department of Interdisciplinary Studies, Faculty of Engineering, Tel
Aviv University, Tel Aviv 69978, Israel}

\begin{abstract}
We consider a system of two discrete nonlinear Schr\"{o}dinger equations,
coupled by nonlinear and linear terms. For various physically relevant
cases, we derive a modulational instability criterion for plane-wave
solutions. We also find and examine domain-wall solutions in the model with
the linear coupling.
\end{abstract}

\maketitle

\section{Introduction}

Modulational instabilities (MIs) have a time-honored history in nonlinear
wave equations. Their occurrences span areas ranging from fluid dynamics
\cite{benjamin67} (where they are usually referred to as the Benjamin-Feir
instability) and nonlinear optics \cite{ostrovskii69,Agrawal} to plasma
physics \cite{taniuti68}.

While earlier manifestations of such instabilities were studied in
continuum systems \cite{Agrawal,hasegawa}, in the last decade the
role of the MI in the dynamics of discrete systems has emerged. In particular, 
the MI was analyzed in the context of the
discrete nonlinear Schr{\"{o}}dinger equation \cite{Kivshar}, a
ubiquitous nonlinear-lattice dynamical model \cite{dnc0,ijmpb}.
More recently, it was studied in the context of weakly interacting
trapped Bose-Einstein condensates (BECs), where the analytical predictions based on the discrete model
\cite{prl2002} were found to be in agreement with the
experiment \cite{njp2003} (see also the recent works \cite{konot1,nicolin,florence}, and recent
 reviews in \cite{abdull,mplb}). Additionally, the
development of \textquotedblleft discrete nonlinear
optics\textquotedblright\ (based on nonlinear waveguide arrays)
has recently provided the first experimental observation of the MI
in the latter class of systems \cite{dnc1}.

An extension of these recent works, which is relevant to both BECs in the
presence of an optical-lattice potential \cite{mandel,bernard,twoc,Panos}
and nonlinear optics in photorefractive crystals \cite{moti}, is the case of
multi-component discrete fields. These can correspond to a mixture of two
different atomic species, or different spin states of the same atom, in BECs
\cite{twocbec}, or to light waves carried by different polarizations or
different wavelengths in optical systems \cite{hudock}.

The corresponding two-component model is based on two coupled discrete
nonlinear Schr\"{o}dinger (DNLS) equations,
\begin{eqnarray}
i\frac{\partial u_{1n}}{\partial t}
&=&-d_{1}(u_{1,n+1}+u_{1,n-1}-2u_{1,n})+(s_{11}|u_{1,n}|^{2}+s_{12}|u_{2,n}|^{2})u_{1,n}+cu_{2,n},
\nonumber \\
i\frac{\partial u_{2,n}}{\partial t}
&=&-d_{2}(u_{2,n+1}+u_{2,n-1}-2u_{2,n})+(s_{12}|u_{1,n}|^{2}+s_{22}|u_{2,n}|^{2})u_{2,n}+cu_{1,n}.
\label{ceq1}
\end{eqnarray}This is a discrete analog of the well-known model describing nonlinear
interactions of the above-mentioned light waves through 
self-phase-modulation and cross-phase-modulation (XPM)
\cite{Agrawal,hasegawa}. In optics, Eqs. (\ref{ceq1}) describe an
array of optical waveguides, the evolution variable being the
propagation distance $z$ (rather than time $t$). The choice of the
nonlinear coefficients in the optical models is limited to the
combinations $s_{11}=s_{22}=3s_{12}/2$ for orthogonal linear
polarizations, and $s_{11}=s_{22}=s_{12}/2$ for circular
polarizations or different carrier wavelengths. In BECs, the
coefficients $s_{jk}$ in Eqs.\ (\ref{ceq1}), are related to the
three scattering lengths $\alpha _{jk}$ which account for
collisions between atoms belonging to the same ($\alpha _{jj}$) or
different ($\alpha _{jk}$, $j\neq k$) species; in that case, $\alpha _{ij}>0$
($\alpha _{{ij}}<0$) corresponds to the repulsive (attractive) interaction between the atoms.
The linear coupling between the components, which is accounted for by the
coefficient $c$ in Eqs. (\ref{ceq1}), is relevant in optics for a case of
circular polarizations in an array of optical fibers with deformed
(non-circular) cores, or for linear polarizations in an array of twisted
fibers \cite{hudock}. On the other hand, in the BECs context, $c$ represents the Rabi frequency of 
transitions between two different spin states in a resonant microwave field
\cite{lincoupling,Panos}.

Stimulated by the experimental relevance of the MI in discrete coupled
systems, the aim of the present work is to develop a systematic study of the
instability in the two-component dynamical lattices. We give an analytical
derivation of the MI criteria for the case of both the nonlinear and linear
coupling between the components. As a result of the analysis, we also find a
novel domain-wall (DW)\ stationary pattern in the case of the linear
coupling. The presentation is structured as follows: In the following
sections we
derive plane-wave solutions and analyze their stability, corroborating it
with a numerical analysis of the stability intervals. We do this for the
model with the linear coupling in section II and for the one with the 
purely nonlinear coupling in section III.
In section IV, we
examine DW states in the linearly coupled lattices. Finally, the results and 
findings are
summarized in section V.

\section{The model with the linear coupling}

We look for plane-wave solutions in the form
\begin{equation}
u_{jn}=A_{j}\exp \left[ i(q_{j}n-\omega _{j}t)\right] ,\quad j=1,2.
\label{plane}
\end{equation}
The linear coupling imposes the restrictions $q_{1}=q_{2}\equiv q$ and $\omega _{1}=\omega _{2}\equiv \omega $. Inserting Eq. (\ref{plane}) into
Eqs. (\ref{ceq1}) yields
\begin{eqnarray}
\omega A_{1} &=&-2d_{1}(\cos {q}-1)A_{1}+(s_{11}A_{1}^{2}+s_{12}A_{2}^{2})A_{1}+cA_{2},  \nonumber \\
\omega A_{2} &=&-2d_{2}(\cos
{q}-1)A_{2}+(s_{12}A_{1}^{2}+s_{22}A_{2}^{2})A_{2}+cA_{1}.
\label{dwnew}
\end{eqnarray}In the particular, but physically relevant, symmetric case, with $d_{1}=d_{2} $ and $s_{11}=s_{22}$, it follows from here that the amplitudes $A_{1,2}$ obey an equation $[(s_{11}-s_{12})A_{1}A_{2}-c]\left(A_{1}^{2}-A_{2}^{2}\right)=0$, hence either of the
following two relations must then be satisfied:
\begin{equation}
A_{1}=\pm A_{2};  \label{plane1}
\end{equation}\begin{equation}
A_{1}A_{2}=\frac{c}{s_{11}-s_{12}}.  \label{plane2}
\end{equation}

In the case of Eq. (\ref{plane1}), a nonzero solution has
$A_{1}=\pm \sqrt{\left( 2d_{1}(\cos {q}-1)+\omega \mp c\right)
/\left( s_{11}+s_{22}\right) }$. It exists with $s_{11}+s_{12}>0$,
provided $2d_{1}(\cos {q}-1)+\omega \mp c>0$, and with
$s_{11}+s_{12}<0$, if $2d_{1}(\cos {q}-1)+\omega \mp c<0$. When
$s_{11}=-s_{12}$, one obtains solutions of the form $\left(
A_{1},A_{2}\right) =\left( \pm A,A\right) $ with arbitrary $A$ and
$-2d_{1}(\cos {q}-1)=\omega \mp c$.

On the other hand, from Eq. (\ref{plane2}), one finds that
\begin{eqnarray}
A_{1}^{2} &=&\frac{\omega +2d_{1}(\cos {q}-1)}{2s_{11}}  \nonumber \\
&&\pm \frac{1}{2}\sqrt{\left( \frac{\omega +2d_{1}(\cos
{q}-1)}{s_{11}}\right) ^{2}-\frac{4c^{2}}{(s_{11}-s_{12})^{2}}}
\label{tip2}
\end{eqnarray}under the restriction that this expression must be positive. 
When $c\neq 0$, solutions of this type exist as long as $\left[
\omega +2d_{1}(\cos {q}-1)\right] s_{11} > 0$ (as the term under the square root 
is smaller in magnitude than the
one outside) and the argument
of the square root in (\ref{tip2}) is non-negative. The first
condition implies $\omega >-2d_{1}(\cos {q}-1)$, for $s_{11}>0$,
and $\omega <-2d_{1}(\cos {q}-1)$, for $s_{11}<0$. In the case $s_{12}=0$,
Eq. (\ref{tip2}) takes the simpler form
\begin{equation}
A_{1}^{2}=\left( 2s_{11}\right) ^{-1}\left[ \omega +2d_{1}(\cos {q}-1)\pm
\sqrt{(\omega +2d_{1}(\cos {q}-1))^{2}-4c^{2}}\right] .
\end{equation}
For both $s_{12}=0$ and $s_{12}=2s_{11}$, 
it is necessary to impose the condition $|\omega +2d_{1}(\cos {q}-1)|~\geq 2c$ for the 
solutions to be real. We note that
these relations are similar to those derived in Ref. \cite{Panos},
where linearly and nonlinearly coupled systems of continuum NLS equation
were considered.

It is also worth noting that the existence of two \textit{distinct}
uniform states with a fixed product from Eq. (\ref{plane2})
suggests a possibility of a domain-wall (DW) solution in the model
with the linear coupling. DW solutions in nonlinearly coupled
discrete nonlinear Schr{\"{o}}dinger equations were examined in
Ref. \cite{dw} (following an analogy with the continuum ones of
Ref. \cite{boris}). However, the present case is different, as
both uniform states have non-vanishing amplitudes in both
components, and, as seen from Eq. (\ref{plane2}), the DWs may
exist only if $c\neq 0$.
 This possibility is examined in more
detail in section IV.
 
To examine the stability of the plane waves, we substitute
\begin{equation}
u_{jn}(x,t)=[A_{j}+B_{jn}(x,t)]\exp [i(qn-\omega t)]  
\label{ceq1a}
\end{equation}
into Eqs. (\ref{ceq1}), to obtain a system of two 
coupled linearized 
equations for the perturbations $B_{j}(x,t)$. 
Furthermore, assuming a general solution of the above-mentioned system of the form
\begin{equation}
B_{jn}=\alpha _{j}\cos (Qn-\Omega t)+i\beta _{j}\sin (Qn-\Omega t),
\label{ceq3}
\end{equation}
where $Q$ and $\Omega $ are the wavenumber and frequency of perturbation,
we arrive at a set of four homogeneous equations
for $\alpha _{1}$, $\beta _{1}$, $\alpha _{2}$ and $\beta _{2}$. The latter  have
a nontrivial solution if $Q$ and $\Omega $ satisfy the dispersion relation
\begin{eqnarray*}
&&\left[ (\Omega -2d_{1}\sin {Q}\sin {q})^{2}-\left( 2d_{1}\cos {q}(\cos {Q}-1)+c\frac{A_{2}}{A_{1}}\right) \left( 2d_{1}\cos {q}(\cos {Q}-1)+c\frac{A_{2}}{A_{1}}-2s_{11}A_{1}^{2}\right) \right] \\
&&\times \left[ (\Omega -2d_{2}\sin {Q}\sin {q})^{2}-\left( 2d_{2}\cos {q}(\cos {Q}-1)+c\frac{A_{1}}{A_{2}}\right) \left( 2d_{2}\cos {q}(\cos {Q}-1)+c\frac{A_{1}}{A_{2}}-2s_{22}A_{2}^{2}\right) \right] \\
&&-2c(2s_{12}A_{1}A_{2}+c)(\Omega -2d_{1}\sin {Q}\sin {q})(\Omega
-2d_{2}\sin {Q}\sin {q})-c^{2}\left( 2d_{1}\cos {q}(\cos {Q}-1)+c\frac{A_{2}}{A_{1}}-2s_{11}A_{1}^{2}\right) \\
&&\times \left( 2d_{2}\cos {q}(\cos
{Q}-1)+c\frac{A_{1}}{A_{2}}-2s_{22}A_{2}^{2}\right)
-(2s_{12}A_{1}A_{2}+c)^{2}
\end{eqnarray*}\begin{equation}
\times \left[ \left( 2d_{1}\cos {q}(\cos {Q}-1)+c\frac{A_{2}}{A_{1}}\right)
\left( 2d_{2}\cos {q}(\cos {Q}-1)+c\frac{A_{1}}{A_{2}}\right) -c^{2}\right]
=0.  
\label{ceq4}
\end{equation}
Note that in the absence of
coupling, i.e., $c=s_{12}=0$, we obtain a known relation \cite{Kivshar}
\begin{equation}
(\Omega -2d_{j}\sin {Q}\sin {q_{j}})^{2}=2d_{j}\cos {q_{j}}(\cos
{Q}-1)\left( 2d_{j}\cos {q_{j}}(\cos
{Q}-1)-2s_{jj}A_{j}^{2}\right),
\end{equation}
which gives the MI condition if the right-hand side becomes negative, i.e.,
\begin{equation}
d_{j}\cos {q_{j}}(2d_{j}\cos {q_{j}}\sin ^{2}{(Q/2)}+s_{jj}A_{j}^{2})<0.
\label{k_r}
\end{equation}

We will now consider the particular case $d_{1}=d_{2}\equiv d$, the general
case being technically tractable but too involved. Then, the dispersion
relation (\ref{ceq4}) takes the from
\begin{equation}
(\Omega -2d\sin {Q}\sin {q})^{4}-(K_{1}+K_{2}+K_{3})(\Omega
-2d\sin {Q}\sin {q})^{2}+K_{1}K_{2}-K_{4}=0,  \label{ceq5}
\end{equation}where
\begin{eqnarray*}
K_{1} &=&\left( 2d\cos {q}(\cos {Q}-1)+c\frac{A_{2}}{A_{1}}\right) \left(
2d\cos {q}(\cos {Q}-1)+c\frac{A_{2}}{A_{1}}-2s_{11}A_{1}^{2}\right) , \\
K_{2} &=&\left( 2d\cos {q}(\cos {Q}-1)+c\frac{A_{1}}{A_{2}}\right) \left(
2d\cos {q}(\cos {Q}-1)+c\frac{A_{1}}{A_{2}}-2s_{22}A_{2}^{2}\right) , \\
K_{3} &=&2c(2s_{12}A_{1}A_{2}+c),
\end{eqnarray*}\begin{eqnarray*}
K_{4} &=&c^{2}\left( 2d\cos {q}(\cos {Q}-1)+c\frac{A_{2}}{A_{1}}-2s_{11}A_{1}^{2}\right) \left( 2d\cos {q}(\cos {Q}-1)+c\frac{A_{1}}{A_{2}}-2s_{22}A_{2}^{2}\right)  \\
&&+(2s_{12}A_{1}A_{2}+c)^{2}\left[ \left( 2d\cos {q}(\cos
{Q}-1)+c\frac{A_{2}}{A_{1}}\right) \left( 2d\cos {q}(\cos
{Q}-1)+c\frac{A_{1}}{A_{2}}\right) -c^{2}\right] .
\end{eqnarray*}
It immediately follows from Eq. (\ref{ceq5}) that, to avoid the MI, both
solutions for $(\Omega -2d\sin {Q}\sin {q})_{1,2}^{2}$ should be positive.
Taking into account the binomial nature of the equation, it is concluded that the spatially homogeneous solution is unstable if either the sum $\Sigma=K_{1}+K_{2}+K_{3}$ or the product $\Pi=K_{1}K_{2}-K_{4}$ of the solutions is negative:
\begin{eqnarray}
K_{1}+K_{2}+K_{3} &<&0;  \label{ceqn1} \\
K_{1}K_{2}-K_{4} &<&0.  \label{ceqn2}
\end{eqnarray}

\begin{figure}[tbp]
\includegraphics[height=3.00in,width=4.00in]{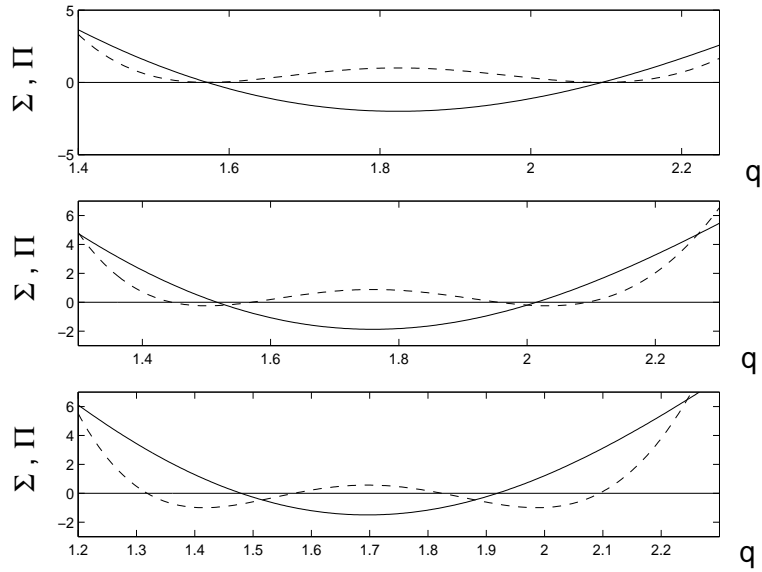} \centering
\caption{The figure shows, for $s_{12}=0$, the cases of $c=0$ (top
panel; unstable for $\protect\pi /2<q<2.0945$), $c=0.25$ (middle
panel; unstable for $1.4455<q<2.0945$) and $c=0.5$ (bottom panel;
unstable for $1.318<q<2.0945$). The solid line shows the sum
$\Sigma $ and the dashed line the product $\Pi $ of the solutions
of Eq. (\protect\ref{ceq5}). The instability takes place in
intervals of the wavenumber $q$ of the unperturbed plane-wave
solution where either $\Sigma $ or $\Pi $ (or both) are negative.}
\label{rfig1}
\end{figure}

\begin{figure}[tbp]
\includegraphics[height=3.00in,width=4.00in]{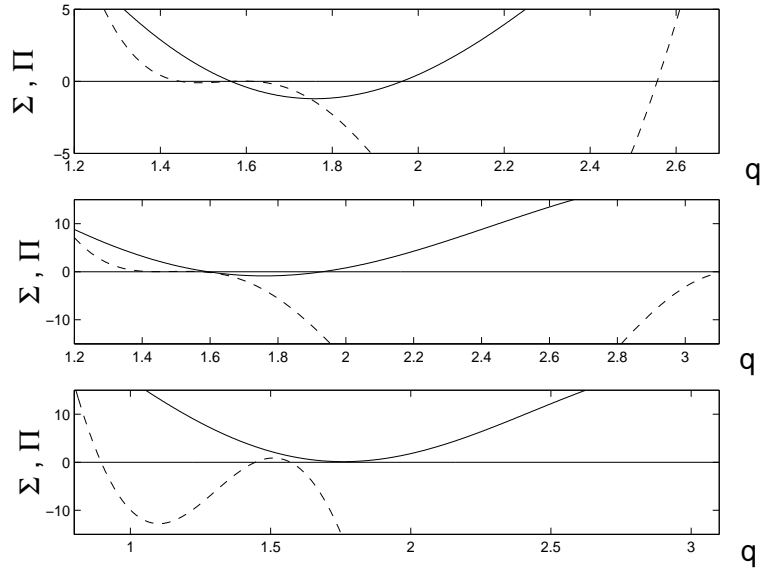} \centering
\caption{Same as Fig. \protect\ref{rfig1} but fixing $c=0.25$ and
varying $s_{12}$. The figure shows the cases of $s_{12}=2/3$ (top
panel; unstable for $1.4455<q<2.556$), $s_{12}=1$ (middle panel;
unstable for $\protect\pi /2<q<\protect\pi $), and $s_{12}=2$
(bottom panel; unstable for $0.8955<q<1.4455$ and $\protect\pi
/2<q<\protect\pi $). } \label{rfig2}
\end{figure}

To proceed further, we may fix the value of the perturbation wavenumber, to investigate 
in what parameter region it would give rise to the MI. 
We illustrate this approach in Figs. \ref{rfig1} and \ref{rfig2},
in which we fix $Q=\pi $ and
$s_{11}=s_{22}=A_{1}=A_{2}=d_{1}=d_{2}=1$ and vary $c$ and
$s_{12}$ (the coefficients of the linear and XPM coupling), to
examine their effect on the stability interval.\ From Eq.
(\ref{k_r}) we see that for these values of the parameters the
modulational unstable region is $\pi /2<q<2\pi /3=2.0945$. It can
be inferred from the figures that $c$ may widen the MI interval by
decreasing its lower edge. On the other hand, $s_{12}$ has a more
complex effect: while making the instability interval larger by
increasing its upper edge (until it reaches $\pi $), it may also
open MI bands within the initially modulationally stable region
(see, e.g., the lower panel of Fig. \ref{rfig2}).

\begin{figure}[tbp]
\includegraphics[height=3.00in,width=3.00in]{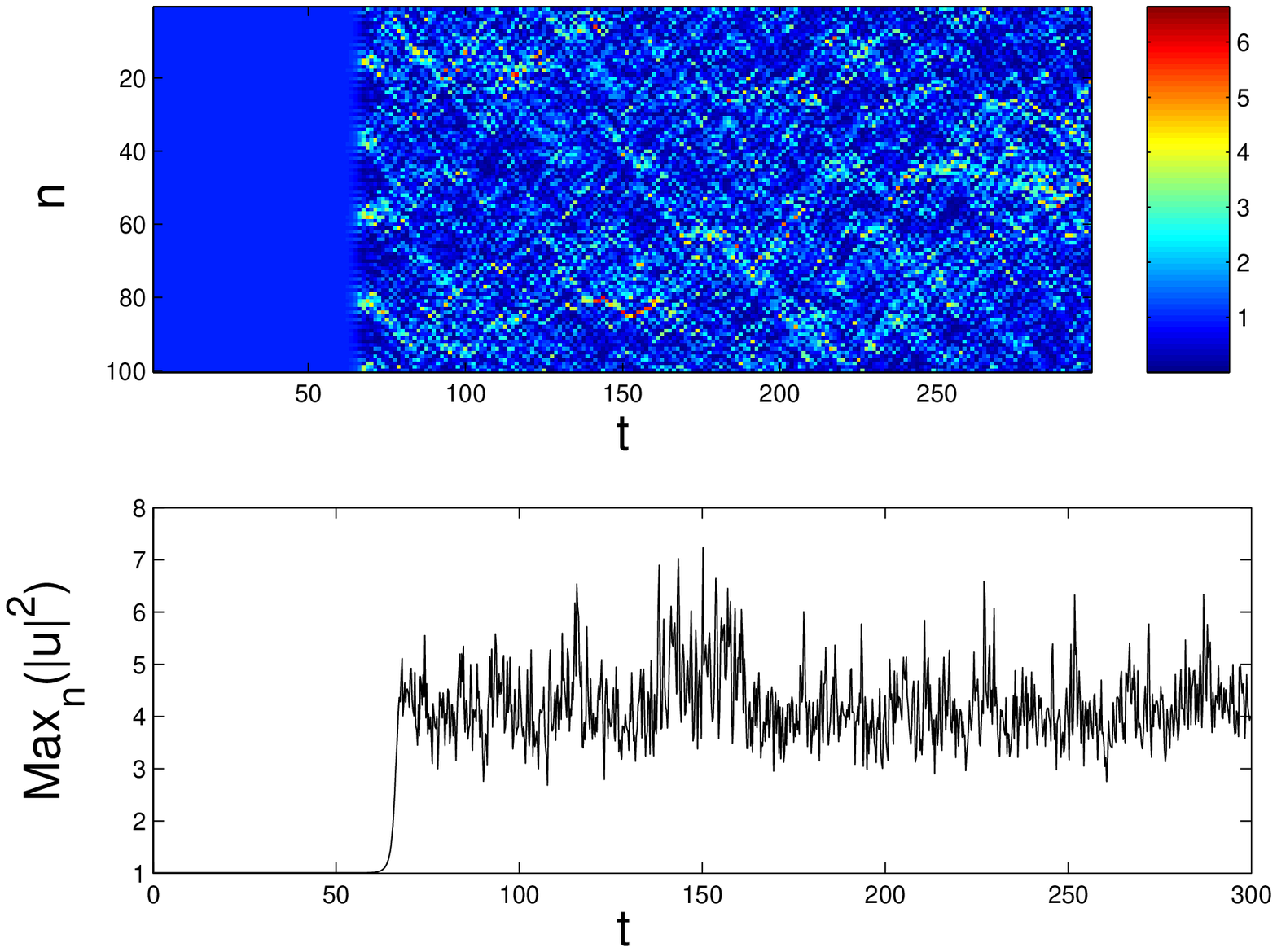} \includegraphics[height=3.00in,width=3.00in]{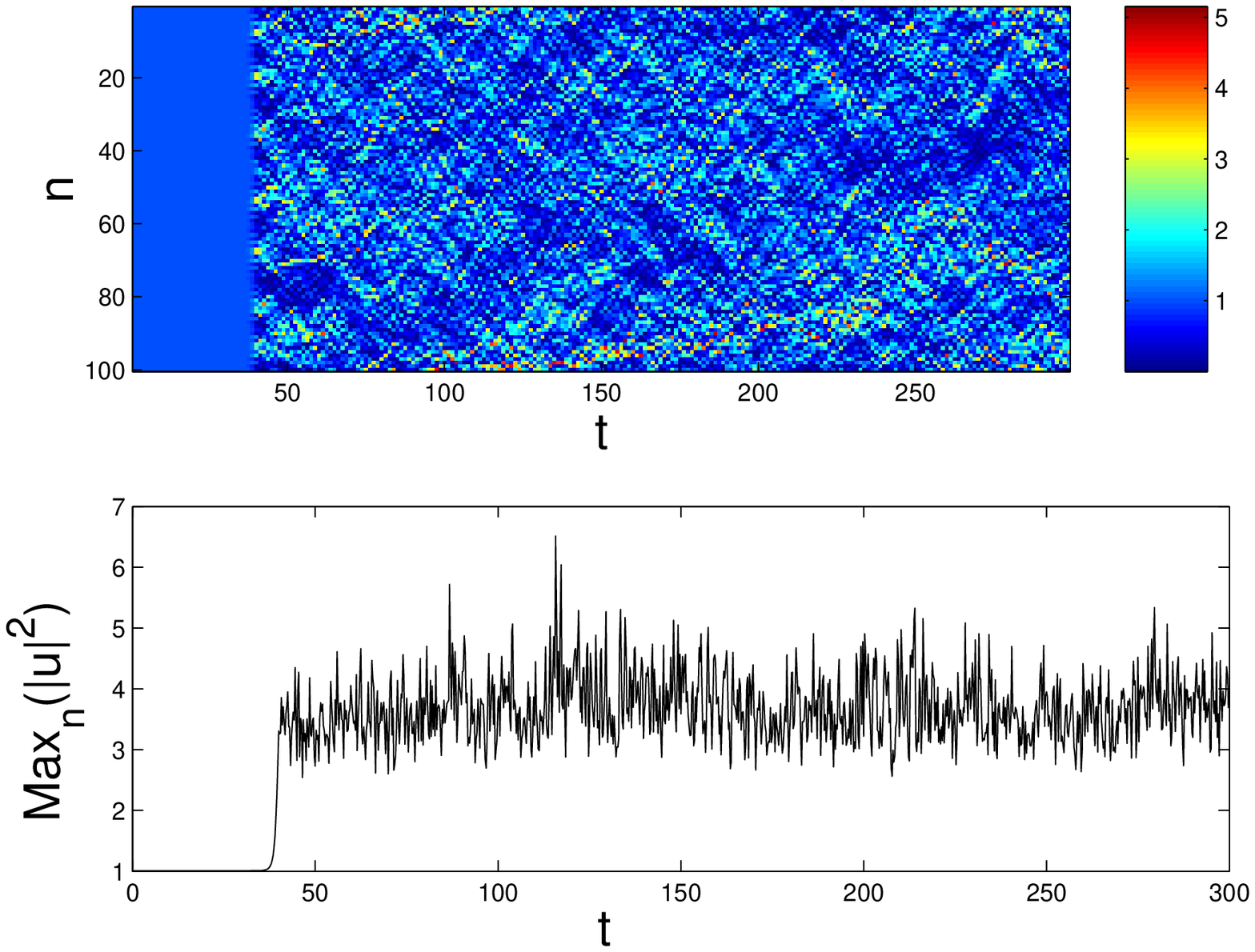} \centering
\caption{Left panels: Spatio-temporal ($n,t$) contour plots of the squared absolute value of one of the two components of the solution (top) and temporal evolution of the maximum of the solution's amplitude (bottom), $c=0.5$, $s_{12}=0$ and $q=\protect\pi /2$. Right panels: Same as the left ones but for $c=0.5$, $s_{12}=2/3$ and $q=\protect\pi /2$.}  
\label{rfig2a}
\end{figure}

For reasons of completeness, we also illustrate nonlinear
development of the MI in the coupled nonlinear lattices in some
typical examples. In particular, the role of the MI in generating
large-amplitude excitations in the presence of the linear coupling
only ($s_{12}=0$) is illustrated in the left panel of Fig.
\ref{rfig2a}, for $c=0.5$. In the right panel of the figure, the
MI in the presence of both linear and nonlinear coupling ($c=0.5$,
$s_{12}=2/3$) is shown. It is readily observed that the nonlinear coupling
enhances the instability growth rate and, hence, the MI sets in earlier.

\section{The case of the purely nonlinear coupling}

A physically relevant case that we also wish to consider here is the one with
only the nonlinear coupling present, i.e., $c=0$. The corresponding
dispersion relations read
\begin{equation}
\omega _{j}=-2d_{j}(\cos {q_{j}}-1)+s_{j1}A_{1}^{2}+s_{j2}A_{2}^{2},\,j=1,2.
\label{ceq7}
\end{equation}

To study the stability of the plane waves in this case, we use Eq.
(\ref{ceq1a}) as before, and obtain the following dispersion
relation for the perturbation wavenumber and frequency
\begin{eqnarray}
&&\left[ (\Omega -2d_{1}\sin {Q}\sin {q_{1}})^{2}-2d_{1}\cos
{q_{1}}(\cos {Q}-1)\left( 2d_{1}\cos {q_{1}}(\cos
{Q}-1)-2s_{11}A_{1}^{2}\right) \right]
\nonumber \\
\times &&\left[ (\Omega -2d_{2}\sin {Q}\sin {q_{2}})^{2}-2d_{2}\cos {q_{2}}(\cos {Q}-1)\left( 2d_{2}\cos {q_{2}}(\cos {Q}-1)-2s_{22}A_{2}^{2}\right) \right]  \nonumber \\
- &&4(2s_{12}A_{1}A_{2})^{2}d_{1}d_{2}\cos {q_{1}}\cos
{q_{2}}(\cos {Q}-1)^{2}=0.  
\label{pureNonlin}
\end{eqnarray}
When $s_{12}=0$, the known result (\ref{k_r}) for the one-component case is 
easily retrieved. If we let $d_{1}=d_{2}\equiv d$ and $q_{1}=q_{2}\equiv q$,
then Eq. (\ref{pureNonlin}) is simplified as follows:
\begin{equation}
(\Omega -2d\sin {Q}\sin {q})^{4}-2K_{5}(\Omega -2d\sin {Q}\sin
{q})^{2}+K_{6}=0,  
\label{ceq9}
\end{equation}
where
\begin{eqnarray}
K_{5} &=&2d\cos {q}(\cos {Q}-1)\left( 2d\cos {q}(\cos {Q}-1)-(s_{11}A_{1}^{2}+s_{22}A_{2}^{2})\right) ,  \nonumber \\
K_{6} &=&(2d\cos {q}(\cos {Q}-1))^{2}\left( (2d\cos {q}(\cos
{Q}-1))^{2}-2(s_{11}A_{1}^{2}+s_{22}A_{2}^{2})2d\cos {q}(\cos
{Q}-1) + 4A_{1}^{2}A_{2}^{2}(s_{11}s_{22}-s_{12}^{2})\right) \nonumber \\
\label{ceq10}
\end{eqnarray}

\begin{figure}[tbp]
\includegraphics[height=3.00in,width=4.00in]{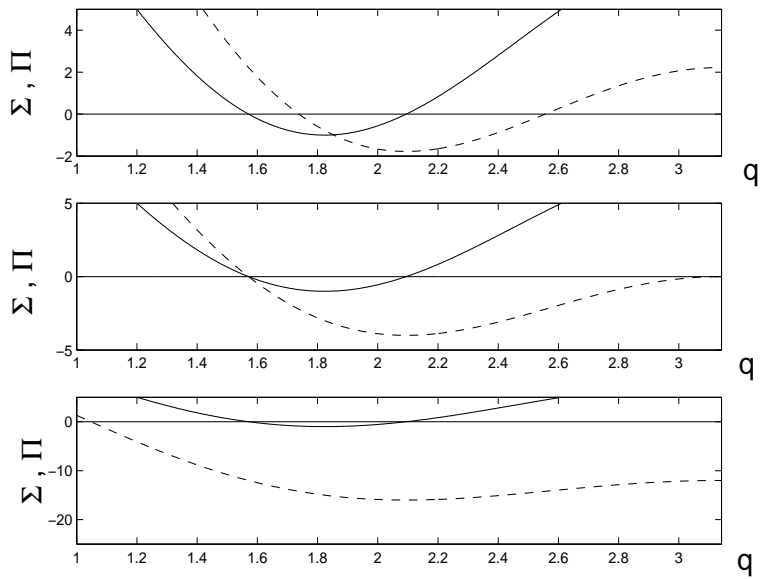} \centering
\caption{Same as Fig. \protect\ref{rfig2} but setting $c=0$ and
varying $s_{12}$. The figure shows the cases of $s_{12}=2/3$ (top
panel; unstable for $\protect\pi /2<q<2.5550$), $s_{12}=1$ (middle
panel; unstable for $\protect\pi /2<q<\protect\pi $), and
$s_{12}=2$ (bottom panel; unstable for $1.0472<q<\protect\pi $).}
\label{rfig3}
\end{figure}

We will follow the lines of the analysis outlined above for the case of the
linear coupling. Examining the product and the sum of the roots of Eq.
(\ref{ceq9}) for $(\Omega -2d\sin {Q}\sin {q})^{2}$, which now read $\Sigma=K_{5}$ and $\Pi=K_{6}$, 
we arrive at the following MI conditions:
\[
K_{5}<0,~\mathrm{or~}K_{6}<0.
\]
The latter can be rewritten, respectively, as:
\begin{eqnarray}
-\frac{1}{2}(s_{11}A_{1}^{2}+s_{22}A_{2}^{2}) &<&2d\cos (q)\sin ^{2}\left(
\frac{Q}{2}\right) <0;  \label{ceqn5} \\
K_{-} &<&-4d\cos (q)\sin ^{2}\left( \frac{Q}{2}\right) <K_{+},  \label{ceqn6}
\end{eqnarray}with $K_{\pm }\equiv s_{11}A_{1}^{2}+s_{22}A_{2}^{2}\pm \sqrt{(s_{11}A_{1}^{2}+s_{22}A_{2}^{2})^{2}-4A_{1}^{2}A_{2}^{2}(s_{11}s_{22}-s_{12}^{2})}$. It is readily observed, in this case as well, that the coupling between the two
components tends to expand the band of the MI wavenumbers with respect to the
single-component case. The effect of the variation of $s_{12}$ in this case,
for fixed $s_{11}=s_{22}=A_{1}=A_{2}=d_{1}=d_{2}=1$ and $Q=\pi $, is shown
in Fig. \ref{rfig3}. Notice also that Eqs. (\ref{ceqn5}) and (\ref{ceqn6}) suggest 
multi-component generalizations of the MI criteria.

\begin{figure}[tbp]
\includegraphics[height=3.00in,width=4.00in]{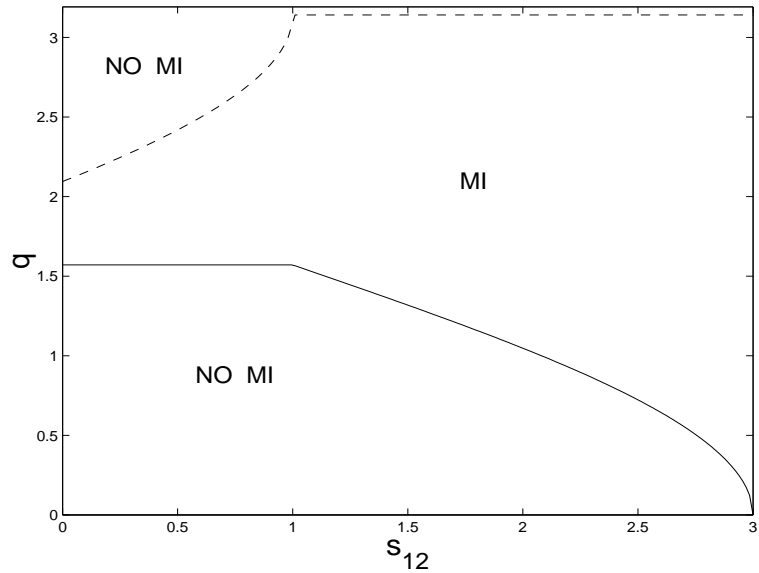} \centering
\caption{The threshold wavenumbers $q$ for the modulational
instability vs. $s_{12}$, as found from Eqs.
(\protect\ref{t1}) and (\protect\ref{t2}) (for the case of $A=d=s=1$).} 
\label{fig_add}
\end{figure}

When $s_{11}=s_{22}\equiv s$, $A_{1}=A_{2}\equiv A$ and $s,s_{12}>0$, the MI
conditions given by Eqs. (\ref{ceqn5})-(\ref{ceqn6}) can be written in a
compact form,
\begin{eqnarray}
-sA^{2} &<&2d\cos (q)\sin ^{2}\left( Q/2\right) <0;  \label{ceqn5a} \\
(s-s_{12})A^{2} &<&-2d\cos (q)\sin ^{2}\left( Q/2\right) <(s+s_{12})A^{2}.
\label{ceqn6a}
\end{eqnarray}
From Eq. (\ref{ceqn5a}), we obtain that
 (for $Q=\pi$)
\begin{equation}
\frac{\pi }{2}<q<\pi -\arccos \left( \frac{sA^{2}}{2d}\right),
\label{t1}
\end{equation}
while similarly from Eq. (\ref{ceqn6a}), it follows that
\begin{equation}
\pi -\arccos \left( \frac{(s-s_{12})A^{2}}{2d}\right) <q<\pi -\arccos \left(
\frac{(s+s_{12})A^{2}}{2d}\right).
\label{t2}
\end{equation}
Examining the latter relations in more detail, we conclude that, for $s>s_{12}$, 
the region of unstable wavenumbers is $\pi /2<q<\pi -\arccos
\left( \frac{(s+s_{12})A^{2}}{2d}\right) $, while, for $s<s_{12}$,
Eq. (\ref{t2}) contains the interval of unstable $q$. This is
illustrated for the case with $A=d=s=1$ as a function of $s_{12}$
in Fig. \ref{fig_add}.

\begin{figure}[tbp]
\includegraphics[height=3.00in,width=4.00in]{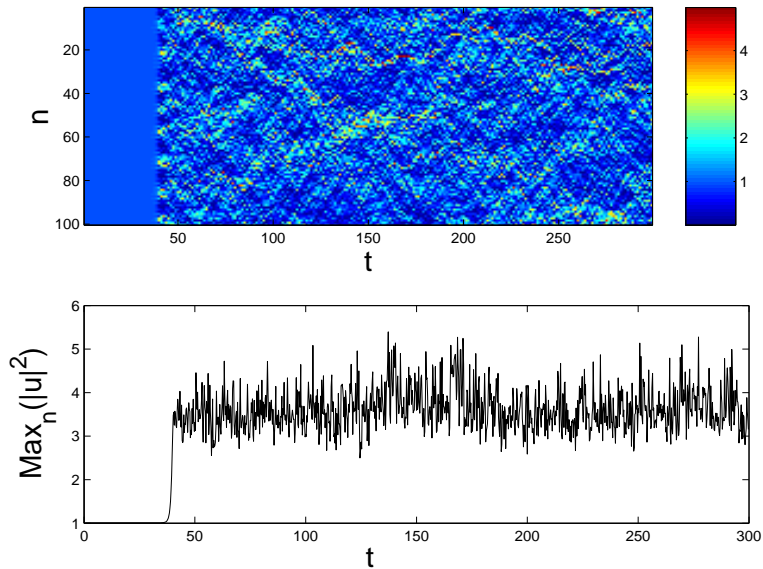} \centering
\caption{Same as Fig. \protect\ref{rfig2a}, but for $s_{12}=2/3$ and $c=0$
(purely nonlinear coupling); $q=\protect\pi /2$.}
\label{rfig3a}
\end{figure}

A typical example of the simulated development of the MI in the case of
purely nonlinear coupling is shown in Fig. \ref{rfig3a} for $s_{12}=2/3$ and
$q=\pi /2$.

\section{Domain walls in the system with the linear coupling}


The expressions (\ref{tip2}) for the amplitudes of the plane-wave 
solutions suggest a novel possibility: 
Focusing more specifically on the so-called anti-continuum limit of $d=0$ 
(we use $d_1=d_2 \equiv d$) and following
the path of \cite{dw}, we can construct \textit{domain-wall} (DW)\ solutions 
that connect the homogeneous state given by Eq. (\ref{tip2}) with its
``conjugate'' state of $A_2=c/(A_1 (s_{11}-s_{12}))$. Such a solution
can then be continued for finite coupling, to examine its spatial
profile and dynamical stability.

\begin{figure}[tbp]
\includegraphics[height=3.00in,width=3.00in]{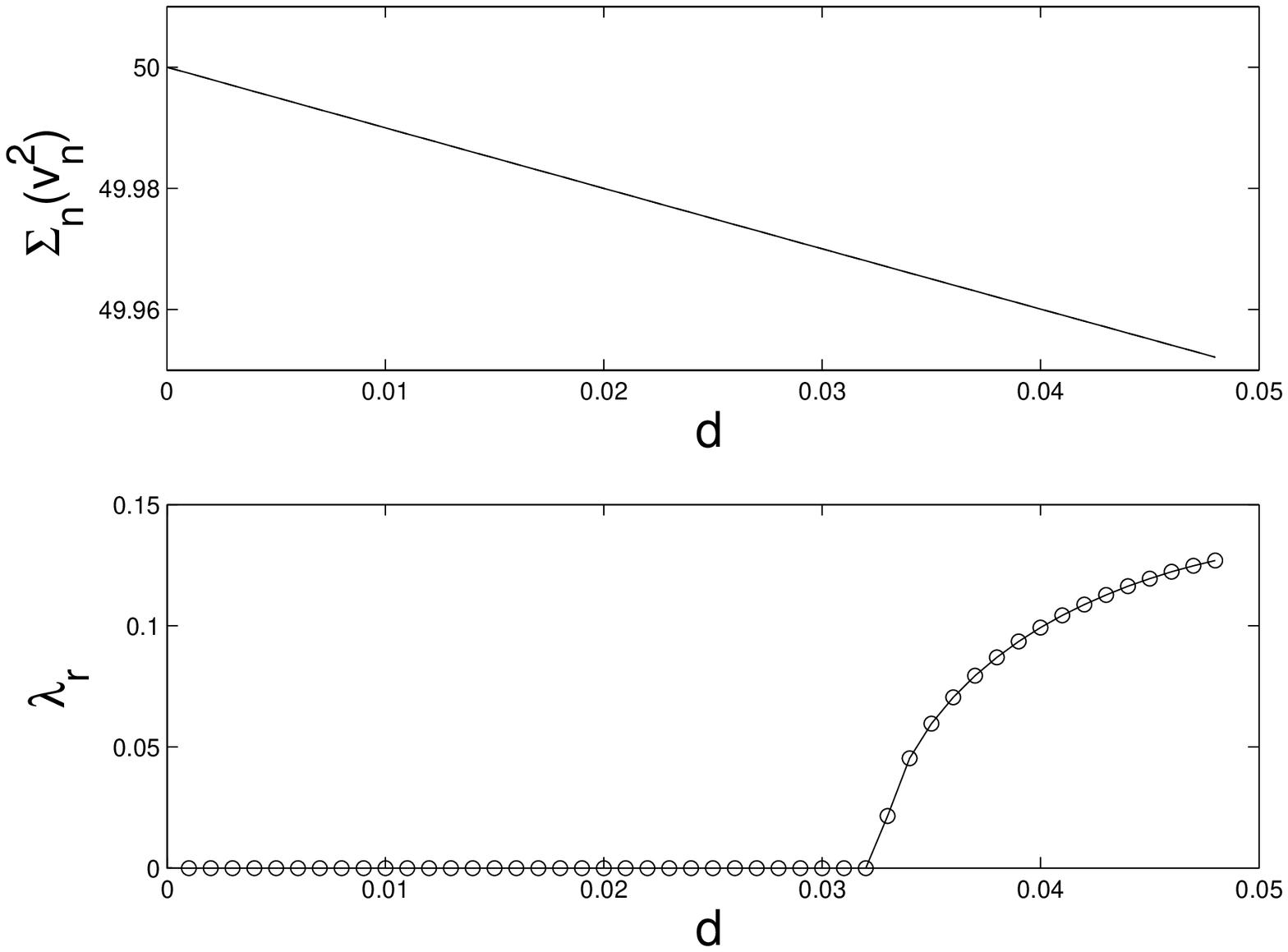} \includegraphics[height=3.00in,width=3.00in]{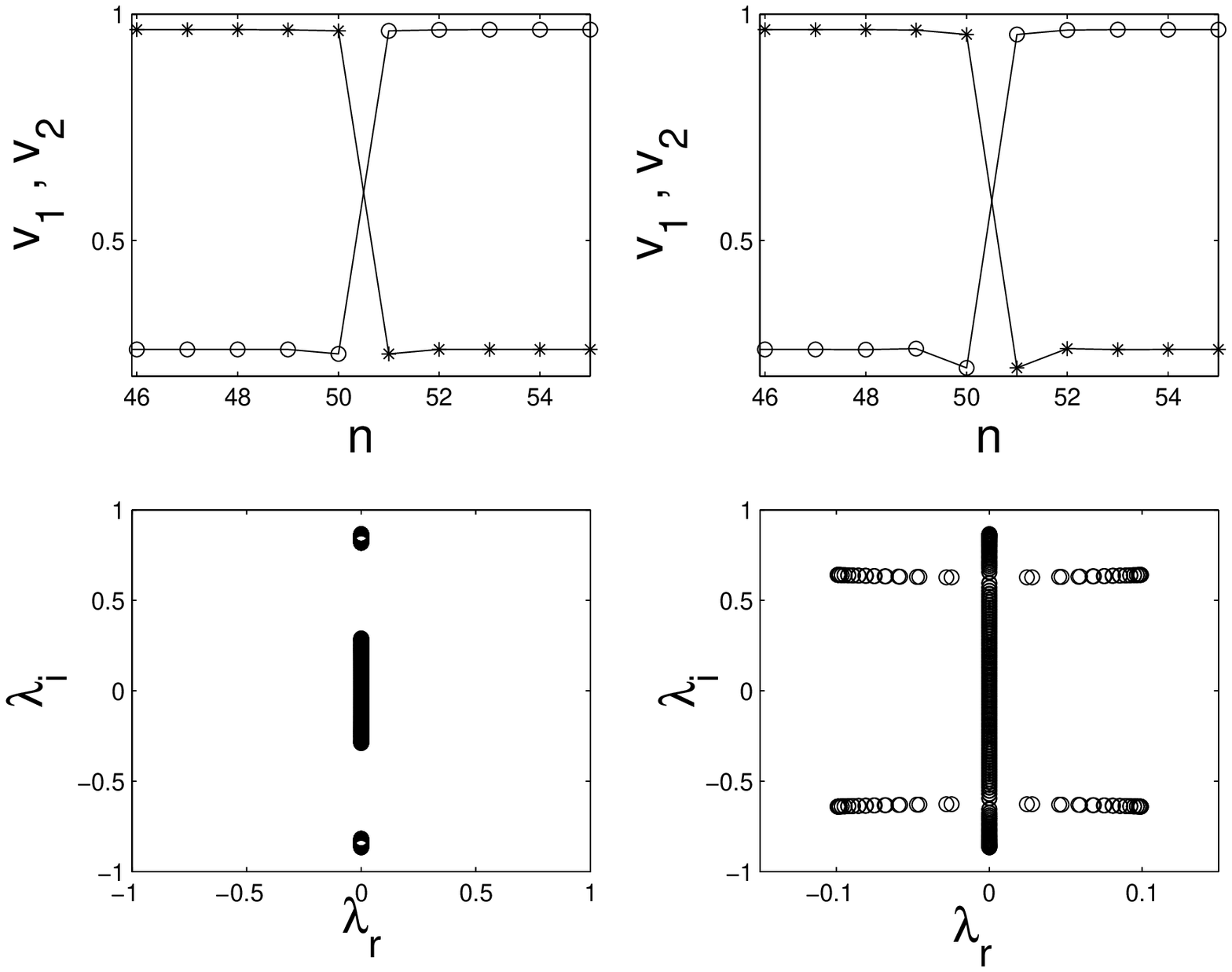} \centering
\caption{The left panel (top subplot) shows the evolution of the
norm of the DW\ solution $u_{jn}=\exp (-i\protect\omega t)v_{jn}$,
as a function of the inter-site coupling constant $d$ using
continuation from the \textit{anti-continuum limit}, $d=0$. The
bottom subplot shows the most unstable eigenvalue of small
perturbations around the solution as a function of $d$. The DW
becomes unstable for $d>0.033$. The profiles of the solution (top
subplots) and the respective linear-stability eigenvalues (bottom
subplots) are shown in the right panel for $d=0.01$ (stable; left
subplots) and $d=0.04$ (unstable; right subplots). The parameters
are $s_{12}=0$, $c=0.25$, $s=\protect\omega =1$.} \label{rfig4}
\end{figure}

An example of such a solution, for the case where the nonlinear coupling is
absent, is given in Fig. \ref{rfig4}. It is observed that the solution is
stable for $d<0.033$, and it becomes unstable due to a cascade of
oscillatory instabilities (through the corresponding eigenvalue quartets)
for larger values of $d$.

\begin{figure}[tbp]
\includegraphics[height=3.00in,width=4.00in]{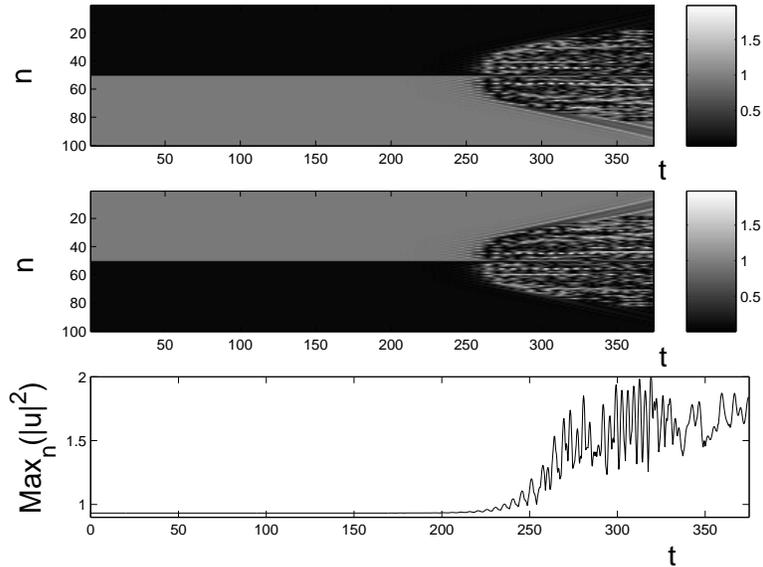}\centering
\caption{The top panel shows a grayscale image of the spatio-temporal
evolution of the intensity of the first field, $\left\vert
u_{1n}(t)\right\vert ^{2}$. The middle panel shows the same for the second
field. The bottom panel displays the time evolution of the amplitude (the
spatial maximum) of the first field.}
\label{rfig4a}
\end{figure}

We have also examined the evolution of the DWs when they are unstable. A
typical result is displayed in Fig. \ref{rfig4a}. It is seen from the bottom
panel, which shows the time evolution of the solution's maximum amplitude, that
growth of the oscillatory instability eventually destroys the configuration,
through \textquotedblleft lattice turbulence\textquotedblright. This
apparently chaotic evolution can be attributed to the mixing of a large number of
unstable eigenmodes. The dynamics remain extremely complex despite the eventual
saturation of the instability.

\begin{figure}[tbp]
\includegraphics[height=3.00in,width=3.00in]{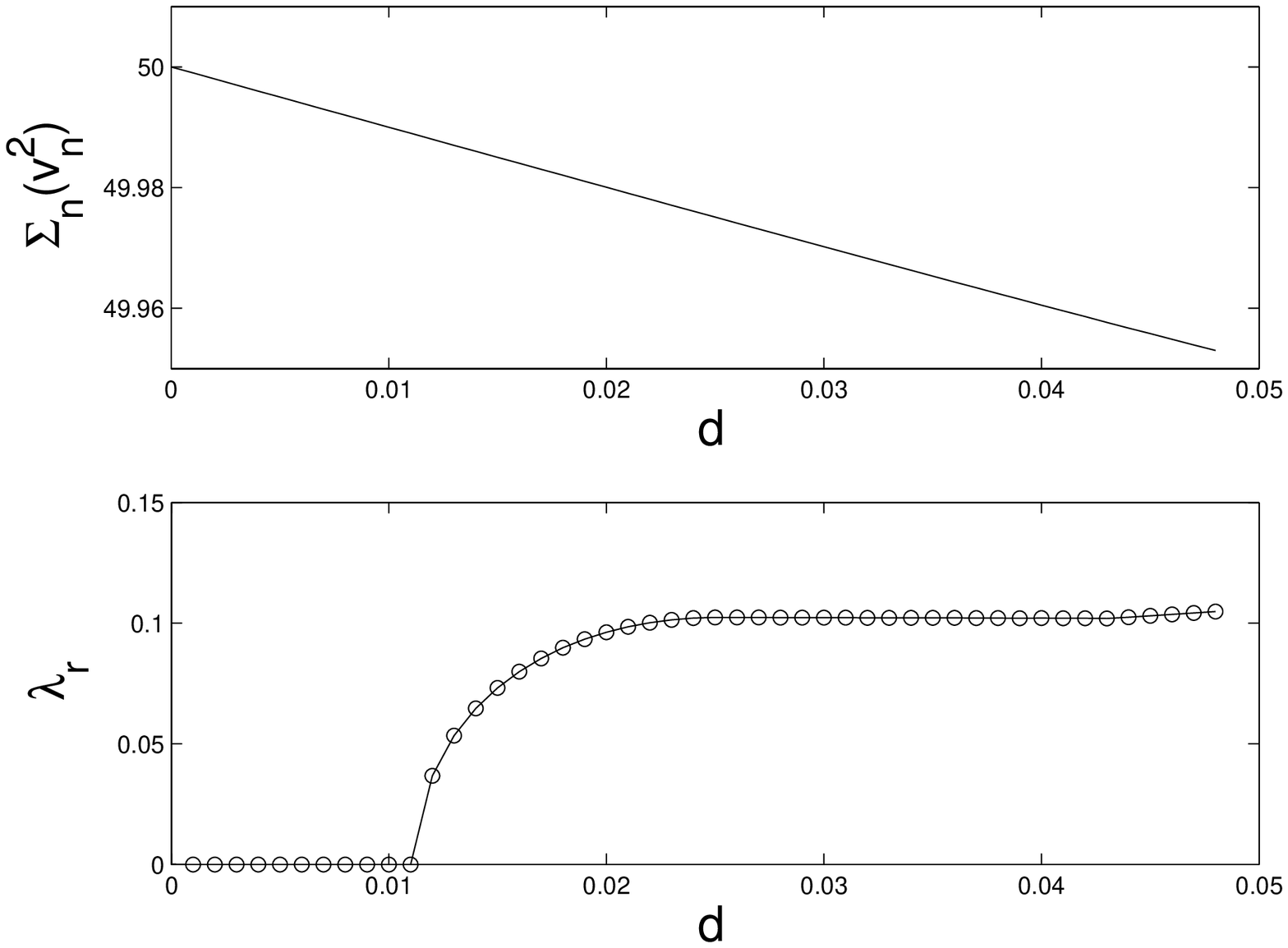} \includegraphics[height=3.00in,width=3.00in]{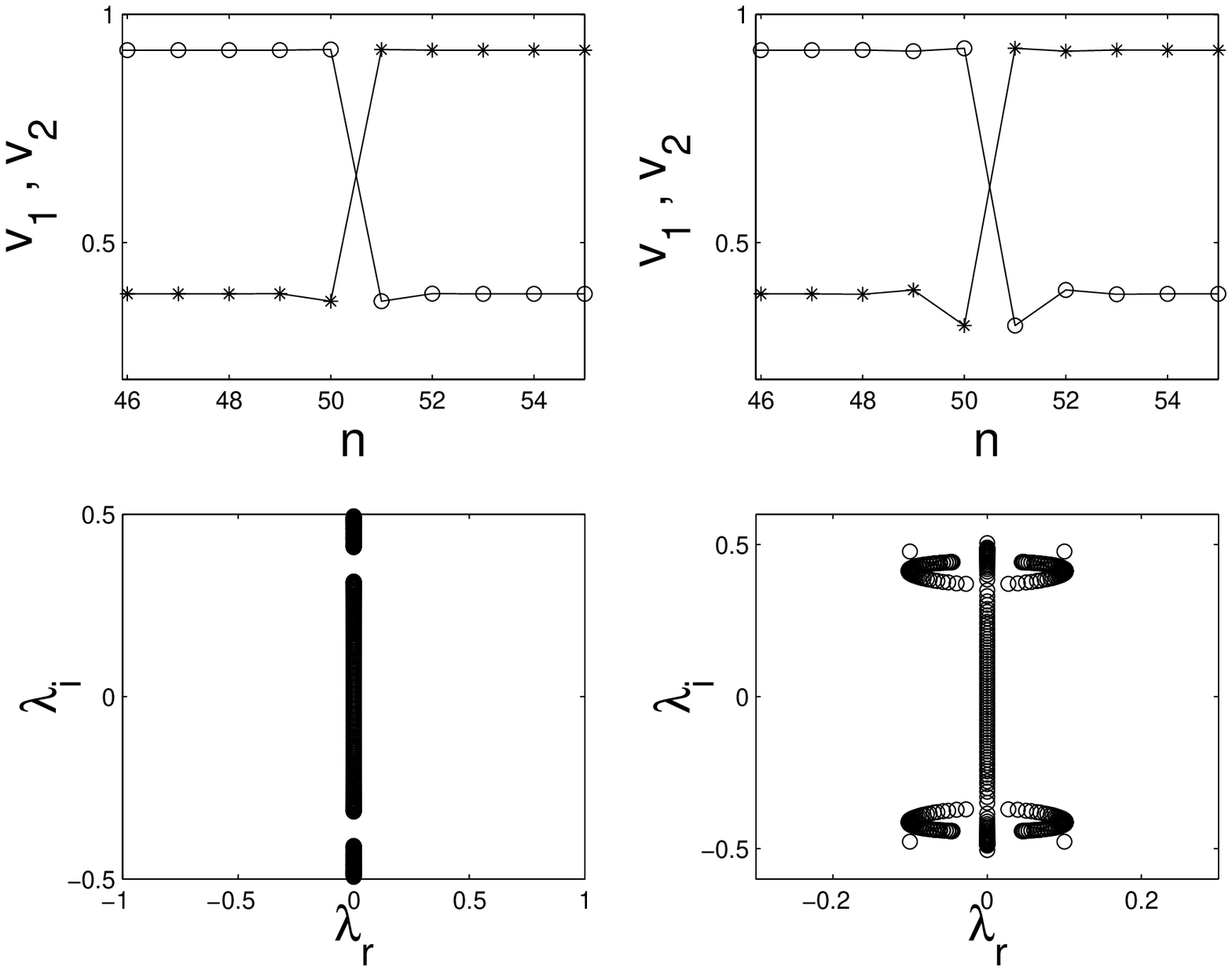} \centering
\caption{Same as Fig. \protect\ref{rfig4}, but for $s_{12}=0.3$. The
instability sets in for $d>0.012$.}
\label{rfig5}
\end{figure}

The increase of the nonlinear coupling constant $s_{12}$ reduces the
stability window of these solutions. In particular, the case of $s_{12}=0.3$
is shown in Fig. \ref{rfig5} in which, the stability window has shrunk to $d<0.012$.

One can also consider a modified DW, where an extra site between the two
domains has equal or opposite amplitudes of the two fields. We have checked
that such solutions are always unstable (due to the presence of real eigenvalue pairs), for
all values of $d$. Still more unstable (with a larger number of unstable
eigenvalues) are more sophisticated DW patterns, with additional
intermediate sites inserted between the two domains.

\section{Conclusions}

In this work, we have examined the extension of the modulational instability
(MI)\ concept to the case of multiple-component discrete fields. We have
shown, in a systematic way, how to formulate the linear MI equations and how to extract the MI criteria. We have also followed the dynamical evolution of the instability by means of
direct simulations, and have identified the effects of the linear and nonlinear
couplings on the range of modulationally unstable wavenumbers. In particular, we have
demonstrated that the joint action of the two couplings may give rise to 
noteworthy features, such as opening of new MI\ bands on the wavenumber 
scale.

Additionally, the identification of a pair of conjugate uniform solutions in
the two-component model has prompted us to examine domain-wall (DW)\
solutions between such states. We were able to demonstrate that the DWs can
be linearly stable, provided that the inter-site coupling in the lattice is
sufficiently weak.

From our results, it is clear that multi-component lattice models have a
rich phenomenology, which is a natural addition to that of single-component
ones. It would be interesting to observe the predicted features in
experimental settings, including weakly coupled BECs and photonic-crystal
nonlinear media.

\textit{Acknowledgements.} ZR and PGK gratefully acknowledge the hospitality
of the Center of Nonlinear Studies of the Los Alamos National Laboratory
where part of this work was performed. PGK also acknowledges the support of
NSF-DMS-0204585, NSF-CAREER and the Eppley Foundation for Research. DJF acknowledges support of the Special Research Account of University of Athens. The work
of BAM was supported, in a part, by the Israel Science Foundation through
the grant No. 8006/03. Research at Los Alamos is performed under the
auspices of the US-DoE.





\begin{thebibliography}{99}
\bibitem{benjamin67} T. B. Benjamin and J. E. Feir, J. Fluid. Mech. \textbf{27}, 417 (1967).

\bibitem{ostrovskii69} L. A. Ostrovskii, Sov. Phys. JETP \textbf{24}, 797
(1969).

\bibitem{Agrawal} G.~P. Agrawal. \newblock {\em Nonlinear Fiber Optics}. \newblock Academic Press, San Diego, CA, 1995.

\bibitem{taniuti68} T. Taniuti and H. Washimi, Phys. Rev. Lett. \textbf{21},
209 (1968); A. Hasegawa, Phys. Rev. Lett. \textbf{24}, 1165 (1970).

\bibitem{hasegawa} A. Hasegawa and Y. Kodama,
\newblock {\it Solitons in
Optical Communications}, Clarendon Press (Oxford 1995)

\bibitem{Kivshar} Yu. S. Kivshar and M. Peyrard, 
Phys. Rev. A \textbf{46}, 3198 (1992).

\bibitem{dnc0} D. N. Christodoulides and R. I. Joseph, 
Opt. Lett. \textbf{13}, 794 (1988)

\bibitem{ijmpb} For recent reviews see, e.g., P. G. Kevrekidis, K. {\O }.
Rasmussen, and A. R. Bishop, Int. J. Mod. Phys. B, \textbf{15}, 2833 
(2001); J. C. Eilbeck and M. Johansson,-- \textit{Proc. of the 3rd Conf.
Localization \& Energy Transfer in Nonlinear Systems (June 17-21 2002, San
Lorenzo de El Escorial Madrid)}, ed. L. V\'{a}zquez \textit{et al.} (World
Scientific, New Jersey, 2003), p. 44 
(arXiv:nlin.PS/0211049).

\bibitem{prl2002} A. Smerzi, A. Trombettoni, P. G. Kevrekidis, and A. R.
Bishop, 
Phys. Rev. Lett. \textbf{89}, 170402, 
(2002)

\bibitem{njp2003} F. S. Cataliotti, L. Fallani, F. Ferlaino, C. Fort, P.
Maddaloni and M. Inguscio, New J. Phys. \textbf{5}, 71 (2003).

\bibitem{konot1} 
V. V. Konotop and M. Salerno 
                 Phys. Rev. A {\bf 65}, 021602 (2002).  

\bibitem{nicolin} M. Machholm, A. Nicolin, C. J. Pethick and H. Smith, Phys. Rev. A {\bf 69} (2004) 043604.

\bibitem{florence} L. Fallani \textit{et al.}, cond-mat/0404045.

\bibitem{abdull} F.Kh. Abdullaev, S.A. Darmanyan and J. Garnier,
Progr. Opt. {\bf 44}, 303 (2002).

\bibitem{mplb} P. G. Kevrekidis and D. J. Frantzeskakis, Mod. Phys. Lett. B
\textbf{18}, 173 (2004).

\bibitem{dnc1} J. Meier \textit{et al.}, Phys. Rev. Lett. \textbf{92},
163902 (2004).

\bibitem{mandel} O. Mandel, M. Greiner, A. Widera, T. Rom, T. W. H\"{a}nsch,
and I. Bloch, Phys. Rev. Lett. \textbf{91}, 010407 (2003).

\bibitem{bernard} B. Deconinck, J. N. Kutz, M. S. Patterson, and B. W.
Warner, J. Phys. A: Math. Gen. \textbf{36}, 5431 (2003).

\bibitem{twoc} P. G. Kevrekidis, G. Theocharis, D. J. Frantzeskakis, B. A.
Malomed and R. Carretero-Gonz\'{a}lez, Eur. Phys. J. D: At. Mol. Opt. Phys.
\textbf{28}, 181 (2004).

\bibitem{lincoupling} R. J. Ballagh, K. Burnett, and T. F. Scott, Phys. Rev.
Lett. \textbf{78}, 1607 (1997); J. Williams, R. Walser, J. Cooper, E.
Cornell, and M. Holland, Phys. Rev. A \textbf{59}, R31 (1999); P. \"{O}hberg
and S. Stenholm, Phys. Rev. A \textbf{59}, 3890 (1999).

\bibitem{Panos} M.A. Porter, P. G. Kevrekidis, and B.A. Malomed,
nlin CD/0401023, Physica D {\bf 196} (2004) 106.

\bibitem{moti} For recent work see, e.g., J. W. Fleischer \textit{et al.},
Phys. Rev. Lett. 92, 123904 (2004); D. N. Neshev \textit{et al.}, Phys. Rev.
Lett. 92, 123903 (2004)

\bibitem{twocbec} See, e.g., C. J.\ Myatt \textit{et al.}, \newblock Phys.\
Rev.\ Lett.\ \textbf{78}, 586 (1997); D. S.\ Hall \textit{et al.},
\newblock Phys.\ Rev.\ Lett.\ \textbf{81}, 1539 (1998); D. M.\
Stamper-Kurn \textit{et al.}, \newblock Phys.\ Rev.\ Lett.\
\textbf{80}, 2027 (1998);
G.\ Modugno \textit{et al.}, \newblock Science \textbf{294}, 1320 (2001);
M.\ Mudrich \textit{et al.}, \newblock Phys.\ Rev.\ Lett.\ \textbf{88},
253001 (2002); M.\ Trippenbach
\textit{et al}., J.\ Phys.\ B \textbf{33}, 4017 (2000); 
S.\ Coen and M.\ Haelterman, Phys.\ Rev.\ Lett.\ \textbf{87}, 140401 (2001);
P. \"{O}hberg and L.\ Santos, \newblock Phys.\ Rev.\ Lett.\ \textbf{86},
2918 (2001); 
Th.\ Busch and J. R.\ Anglin, \newblock Phys.\ Rev.\ Lett.\ \textbf{87},
010401 (2001).

\bibitem{hudock} For a recent discussion of the various optical applications
see, e.g., J. Hudock \textit{et al.},
Phys. Rev. E 67, 056618 (2003).

\bibitem{dw} P. G. Kevrekidis \textit{et al.}, Phys. Rev. E \textbf{67},
036614 (2003).

\bibitem{boris} B.A. Malomed, Phys. Rev. E \textbf{50}, 1565 (1994).
\end{thebibliography}
\end{document}